\documentclass[10pt,conference]{IEEEtran}

\usepackage{cite}
\usepackage{amsmath,amssymb,amsfonts}
\usepackage{algorithmic}
\usepackage{graphicx}
\usepackage{textcomp}
\usepackage{xcolor}
\usepackage{tcolorbox}
\usepackage[T1]{fontenc}
\usepackage{hyperref}

\def\BibTeX{{\rm B\kern-.05em{\sc i\kern-.025em b}\kern-.08em
    T\kern-.1667em\lower.7ex\hbox{E}\kern-.125emX}}

\begin{document}

\title{Extractive Multi Product-Line Engineering}

\author{\IEEEauthorblockN{Kamil Rosiak}
\IEEEauthorblockA{\textit{Technische Universität Braunschweig}\\
Braunschweig, Germany\\
k.rosiak@tu-bs.de}
}

\maketitle

\begin{abstract}
Cloning is a general approach to create new functionality within variants as well as new system variants.
It is a fast, flexible, intuitive, and economical approach to evolve systems in the short run.
However, in the long run, the maintenance effort increases.
A common solution to this problem is the extraction of a product line from a set of cloned variants.
This process requires a detailed analysis of variants to extract variability information. 
However, clones within a variant are usually not considered in the process, but are also a cause for unsustainable software.
This thesis proposes an extractive multi product-line engineering approach to re-establish the sustainable development of software variants.
We propose an approach to re-engineer intra-system and inter-system clones into reusable, configurable components stored in an integrated platform and synthesize a matching multi-layer feature model.
\end{abstract}


\vspace{-3mm}

\section{Introduction}

With an increasing demand for custom-tailored software systems, ad-hoc reuse strategies are often used~\cite{mayrand1996experiment}. 
On the one hand, system functionality is reused within a variant, known as intra-system cloning~\cite{koschke2012large}.
On the other hand, entire variants are copied and adapted for new requirements, also known as clone-and-own or inter-system cloning~\cite{schmid2002economic,fischer2014enhancing,lapena2016towards}.
While code cloning comes with practical benefits~\cite{kapser2008cloning, aversano2007clones, krinke2007study}, research has shown that, with a growing number of clones, they become a significant source of faults and cause problems during development, evolution, and maintenance, unless special care is taken to identify and track existing code clones and their evolution~\cite{roy2007survey,koschke2007survey,juergens2009code, gode2011clone,hotta2010duplicate}.
Development and maintenance costs increase with the number of variants~\cite{rubin2012managing}.
The extraction of a product line is a common approach to solve this problem and a relevant industrial case~\cite{berger2013survey, schmid2002economic,dubinsky2013exploratory}.
The state of the art of extractive software product-line adoption and reverse-engineering deals with clones at the granularity of entire system variants~\cite{krueger2002eliminating, kruger2020promote,martinez2017bottom,martinez2017espla,fischer2015ecco,wille2017extractive}.

However, we argue that clones can appear at any level of granularity.
At the coarsest level of entire variants, a set of cloned variants can be integrated into a software product line. 
Clones of finer granularity within variants can also be extracted into configurable components.
Thus, they can be considered as sub-product lines, which can again have further sub-product lines of their own at even finer granularity levels.
Generalizing this concept, the detection and integration of clones into configurable components at any granularity level can be considered as the extraction of a multi product-line~\cite{holl2012systematic}.

This thesis aims to develop an extractive multi product-line engineering approach to consolidate intra- and inter-system clones to reestablish sustainable development and maintenance of variants.
We propose an automatic refactoring of system variants into an integrated platform, configured with a multi-layer feature model to derive system variants and include configurable components.

Extractive multi product-line engineering enables systems to be re-engineered and transitioned from ad-hoc reuse with clone-and-own to multi product-line engineering with an integrated platform~\cite{antkiewicz2014flexible}.

\vspace{-1mm}
\section{Proposed Approach}

\begin{figure*}[t]
	\includegraphics[width=1.0\linewidth]{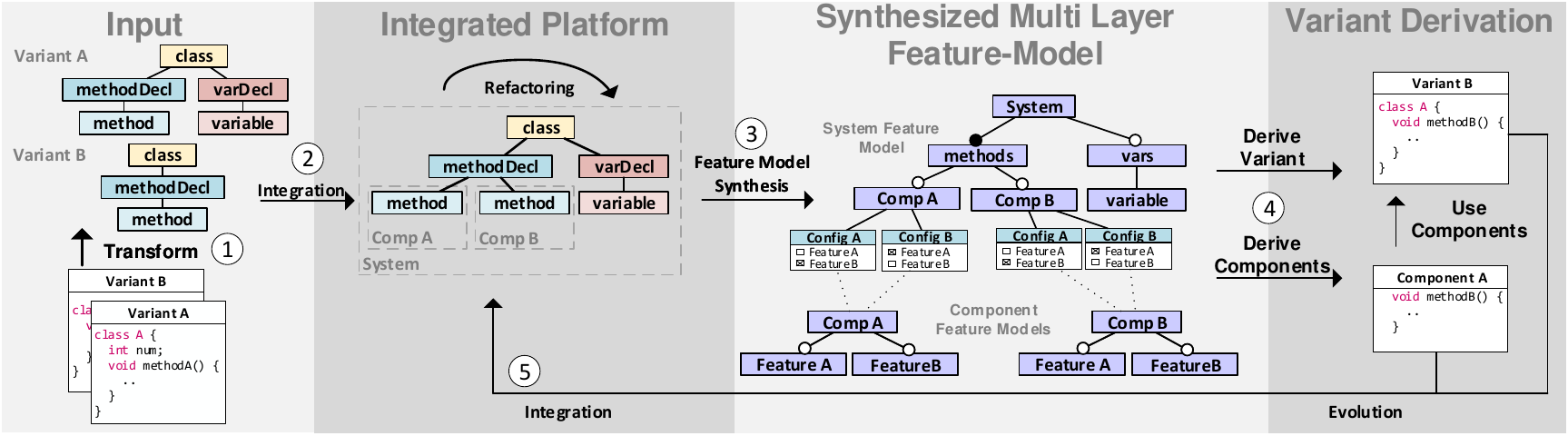}
	\vspace{-4mm}
	\caption{Overview of the extractive multi-product line engineering approach.}
	\vspace{-4mm}
	\label{fig:overview}
\end{figure*}

We propose a technique for extracting multi product-lines from implementation artifacts containing clones at different levels of granularity. 
This approach is applicable to implementation artifacts of different types, such as Java~\cite{arnold2000java} or \mbox{IEC~61131-3}~\cite{tiegelkamp1995iec}.
\autoref{fig:overview} shows an overview of the proposed multi-product line extraction approach using Java as artifact type, which we explain in detail from left to right.
First, artifacts are parsed into a generic graph structure \textcircled{1}.
Artifact adapters translate the respective types of artifacts into the graph structure.
At least one existing variant is required to build the integrated platform \textcircled{2}.
An n-way clone detection approach is applied to the graph structure to identify intra-clones.
Cloned artifacts are identified and refactored into configurable components automatically.
Components are integrated variant-wise into an integrated platform.
We use a pair-wise, iterative, order-invariant variability-mining technique to detect commonalities and differences between already stored artifacts and new variants to be integrated.
The integrated platform can be refactored and annotated with feature information.
Based on the variability and feature information stored in the integrated platform, we generate a fully constrained multi-layer feature model, representing the software product-line of the input system \textcircled{3}.
The integrated platform allows deriving system variants and components \textcircled{4}.
They can be used to evolve system variants as well as to fix bugs across all variants.
This allows sustainable reuse of code within and between variants.
Evolved components and system variants are integrated in the same way as described above \textcircled{5}.

\vspace{-1mm}
\section{Planned Contributions}

The specific contributions of this work are the following:
\textbf{Extractive Multi Product-Line Engineering:}
Krueger~\cite{krueger1992software}, Almeida~\cite{de2005survey} as well as Frankes and Kang~\cite{frakes2005software} provide an overview of software reuse and strategies for the adoption of software. 
However, traditional extractive product-line engineering and clone detection approaches only focus on either inter-system or intra-system clones.
To bridge the gap and reestablish sustainable development, we propose a conceptual process to re-engineer every granularity level of cloned artifacts into a multi product-line.
\\
\textbf{Clone Detection for Any Clone Granularity:}
Research has shown approaches for the detection of clones for different types of artifacts, e.g, trees~\cite{jiang2007deckard,koschke2006clone}, text~\cite{roy2008nicad,ducasse1999language} and tokens~\cite{kamiya2002ccfinder,basit2007efficient}, but also hybrid approachs that use a mix of those artifacts~\cite{kim2011mecc}.
Bellon et al.~\cite{baxter1998clone} showed an approach on abstract syntax trees that can detect exact and near-miss clones.
Göde et al.~\cite{gode2009incremental} developed an incremental clone detection approach based on suffix trees that can analyze software revisions.
However, independent of the technique applied to the subject system to identify clones, most research focuses either on the detection of intra-system or inter-system clones~\cite{Svajlenko2015}.
We propose a clone detection approach, which allows detecting clones at any granularity level.
Moreover, we show an automatic refactoring approach, which allows migrating clones into configurable components.
\\
\textbf{Merging Approach:}
After differencing a set of artifacts, they are post-processed, and an output is generated.
Rubin et al.~\cite{rubin2013n,rubin2013managing} proposed an n-way model merge approach to merge multiple input models into one.
Other research used an iterative pair-wise merge for block-based languages~\cite{wille2014managing,wille2016custom}.
However, existing n-way merging approaches require a comparison of all variants at once, which means that the matching must be recomputed for every new variant.
State of the art pair-wise merging with iteration is not order-invariant. 
In order to achieve ideal results, the correct order between variants needs to be known.
First, we contribute a taxonomy mining approach to determine the order in which variants should be merged to improve the iterative pair-wise merge approach's quality.
Second, we propose an iterative pair-wise order-invariant model merge approach for cases where not all variants are available upfront.
\\ 
\textbf{Feature Model Synthesis:}
Extracting variability information from variants is a commonly used approach to generate a software product line. 
Schlie et.al.~\cite{schlie2020incremental} proposed incremental feature model synthesis for clone-and-own software systems in models.
Their synthesis only considers feature model elements without cross-tree constraints.
We generalize this approach based on an annotation language to allow feature model synthesis on any artifact type and extend it to consider cross-tree constraints.

\vspace{-1.2mm}
\section{Evaluation Plan}

For the evaluation, we derived the following main research questions:
\vspace{-1mm}
\begin{tcolorbox}
	\textbf{RQ1:} Is our approach for re-engineering of clones of any granularity into a multi product-line \emph{correct}, \emph{scalable}, and \emph{useful}? \\
	\textbf{RQ2:} Is our iterative pair-wise order-invariant model merge approach \emph{correct} and \emph{scalable}?\\
	\textbf{RQ3:} Is our synthesized constrained feature model \emph{correct} and \emph{useful}?
\end{tcolorbox}

Lack of case studies with existing ground truth often limits the evaluation of clone and variability analysis approaches.
To close the gap, we plan to develop a seed-based synthetic clone data set generator that simulates cloning on varying granularity levels and tracks the ground truth.
This technique will allow a precise analysis of precision and recall of our clone detection approach.
This generator is based on the automatic mutation framework for evaluating code clone detection tools from Roy et al.~\cite{roy2009mutation} and extends it with cloning of variants.
We plan to evaluate our approach based on synthetically created scenarios, which allows measuring precision and recall of our clone detection approach for arbitrary granularity level.
Those are key measures to show the correctness of clone detection approaches~\cite{roy2009mutation}.
In a second step, we aim to establish software product-lines from known variant-rich case studies such as the pick and place unit (PPU)~\cite{vogel2014researching} or ArgoUML~\cite{zhang2006argouml}.
For evaluation of the multi product-line extraction, we derive as a base-line all integrated variants and compare if all variants could be derived correctly and the number of required artifacts.
To determine the performance and scalability of our approach, we measure run-time and memory consumption.
To determine our feature model synthesis usability, we measure how many operations need to be applied manually and how usable our approach is.
Moreover, we plan a user study with practitioners from local industry to determine our approach's usefulness in real-world scenarios.

\bibliographystyle{IEEEtran}

\bibliography{bib}

\end{document}